\documentclass[prb,twocolumn,amsmath,showpacs]{revtex4}

\usepackage{graphicx}

\begin{document}
\input epsf

\title {Equilibrium magnetization in the vicinity of the first order 
phase transition in the mixed state of high-$T_c$ superconductors.}

\author {I. L. Landau$^{1,2}$ H. R. Ott$^{1}$}
\affiliation{$^{1}$Laboratorium f\"ur Festk\"orperphysik, ETH 
H\"onggerberg, CH-8093 Z\"urich, Switzerland}
\affiliation{$^{2}$Kapitza Institute for Physical Problems, 117334 
Moscow, Russia}

\date{\today}

\begin{abstract}

	We present the results of a scaling analysis of isothermal 
	magnetization $M(H)$ curves measured in the mixed state of high-$T_c$ 
	superconductors in the vicinity of the established first order phase 
	transition. The most surprising result of our analysis is that the 
	difference $\Delta M$ between the magnetization above and below the 
	transition may have either sign, depending on the particular chosen 
	sample. We argue that this observation, based on $M(H)$ data available 
	in the literature, is inconsistent with the interpretation that the 
	well known first order phase transition in the mixed state of 
	high-$T_c$ superconductors always represents the melting transition 
	in the vortex system.
	
\end{abstract}

\maketitle

Measurements of the magnetization of a type-II superconductor in the 
mixed state provide valuable information about different parameters 
characterizing the superconducting state of the investigated material. 
The irreversible magnetization reflects the pinning strength of vortices, 
and by analyzing the reversible magnetization, different equilibrium 
parameters of the superconducting material, such as critical magnetic 
fields and characteristic lengths may be evaluated. This is why 
magnetization measurements are often used to investigate conventional and 
unconventional superconductors. For instance, the well-known first-order 
phase transition in the mixed state of high-$T_c$ superconductors (HTSC), 
which is usually attributed to the melting of the vortex lattice, was 
discovered and confirmed by magnetization measurements. \cite{2,3,4} In 
this work we apply a recently developed scaling procedure \cite{5} for 
an analysis of $M(H)$ curves above and below this phase transition and 
we find inconsistencies that indicate that the transition is not always 
related to a melting of the vortex lattice. 

Our scaling procedure (see Refs. \onlinecite{5,6,7,8}) is based on the 
fact that, if the Ginzburg-Landau parameter $\kappa$ is temperature 
independent, the magnetic susceptibility $\chi$ in the mixed state of a 
type-II superconductor may be written as 
\begin{equation}
\chi(H,T) = \chi(H/H_{c2}),
\end{equation}
where $H_{c2} = H_{c2}(T)$ is the upper critical field. According to Eq. 
(1), the temperature dependence of $\chi$ is solely determined by the 
temperature variation of $H_{c2}$. Eq. (1) is sufficient to derive a 
relation between the magnetizations $M$ at two different temperatures 
$T$ and $T_0$, which may be written as 
\begin{equation}
M(H/h_{c2},T_0) = M(H,T)/h_{c2}
\end{equation}
where $h_{c2}(T)  = H_{c2}(T)/H_{c2}(T_0)$ is the ratio of the upper 
critical fields at $T$ and $T_0$. This equation is valid if the 
diamagnetic response of the mixed state is the only significant 
contribution to the sample magnetization. It is well known, however, that 
many superconducting materials, including the HTSC's that we consider in 
this work, exhibit sizable paramagnetic susceptibilities in the normal 
state. In order to account for this additional contribution to the sample 
magnetization, the following modification of Eq. (2) needs to be made 
\cite{5} 
\begin{equation}
M(H/h_{c2},T_0) = M(H,T)/h_{c2} + c_0(T)H.
\end{equation}

As is discussed detail in Ref. \onlinecite{5}, Eq. (3) may be used for 
the scaling of equilibrium magnetization $M(H)$ curves measured at 
different temperatures, to obtain $M(H,T_0)$. The scaling parameters 
$h_{c2}(T)$ and $c_0(T)$ are determined by the condition that the 
$M(H,T_0)$ curves, calculated from the magnetization data measured at 
different temperatures, collapse onto the same master curve. In this way 
the temperature dependence of the normalized upper critical field 
$h_{c2}(T)$ and the equilibrium magnetization curve $M_{eff}(H) = 
M(H,T_0)$ are obtained. While in Refs. \onlinecite{5,6,7,8}, the main 
goal was to establish the $h_{c2}(T)$ curves, in the present work we analyze 
the scaled $M(H)$ curves in the vicinity of the first order phase 
transition. 

In our previous work we discussed in detail the conditions under which 
our scaling procedure is valid and we showed that these conditions 
are considerably less restrictive than those that were chosen in 
previously presented analyses of $M(H,T)$ curves in the mixed state 
of HTSC's. \cite{5} Our only assumption, the validity of Eq. (1), 
is rather general and the scaling procedure can be used for the 
analysis of the equilibrium magnetization data independent of the 
sample geometry, the pairing type, or of the particular configuration 
of the mixed state. 

Because the validity Eq. (3) is restricted to equilibrium 
magnetization, only $M(H,T)$ curves collected above the 
irreversibility line can be used for the evaluation of the scaling 
parameters $h_{c2}(T)$ and $c_0(T)$. Nevertheless, some additional 
information may also be gained from the analysis of scaled $M_{eff}(H)$ 
curves calculated from magnetization data below the irreversibility 
line. It was argued in Ref. \onlinecite{14} that, if the calculated 
$M_{eff}(H)$ curves collapse onto a single curve also below the 
irreversibility line, this may be regarded as strong evidence that 
the corresponding branches of the measured $M(H,T)$ curves represent 
the equilibrium magnetization. It was indeed demonstrated in Ref. 
\onlinecite{14} that for many Bi-based HTSC's, the $M(H)$ data 
collected in increasing magnetic fields follow the equilibrium 
magnetization curve $M_{eq}(H)$ even in fields well below the 
irreversibility line. Below we show that this is also true for an 
optimally doped YBa$_2$Cu$_3$O$_{7-x}$ (Y-123) sample, while in 
La$_2$CuO$_{4}$ (La-214) based cuprates, this effect is practically 
unobservable and both branches of the measured $M(H)$ curves deviate 
from $M_{eq}(H)$ just below the irreversibility line. 

We apply the scaling described by Eq. (3) to results of 
magnetization measurements in the vicinity of the first order phase 
transition that are available in the literature. It turns out that 
some of the results of our analysis are in contradiction with the 
vortex-lattice-melting hypothesis. For this reason, we use the more 
general nomenclature of "high-field" and "low-field phase", instead 
of the commonly used notations of vortex liquid and vortex solid.

\begin{figure}[h]
 \begin{center}
  \epsfxsize=0.95\columnwidth \epsfbox {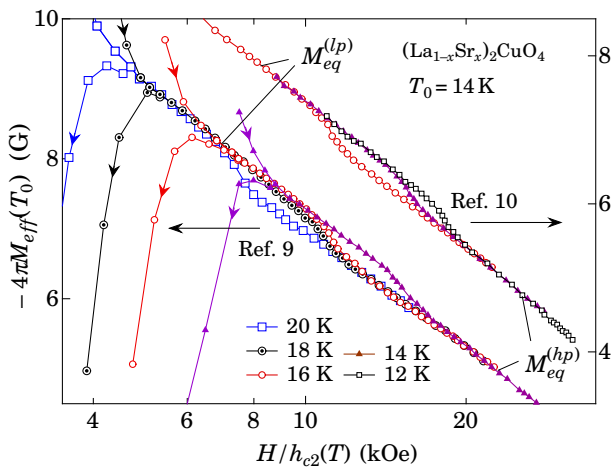}
  \caption{The $M_{eff}(H/h_{c2})$ curves for two La-214 samples 
           studied in Refs. \onlinecite{15} and \onlinecite{16}. For one 
           of the samples, only reversible magnetization data are shown.}
 \end{center}
\end{figure}
\begin{figure}[h]
 \begin{center}
  \epsfxsize=0.95\columnwidth \epsfbox {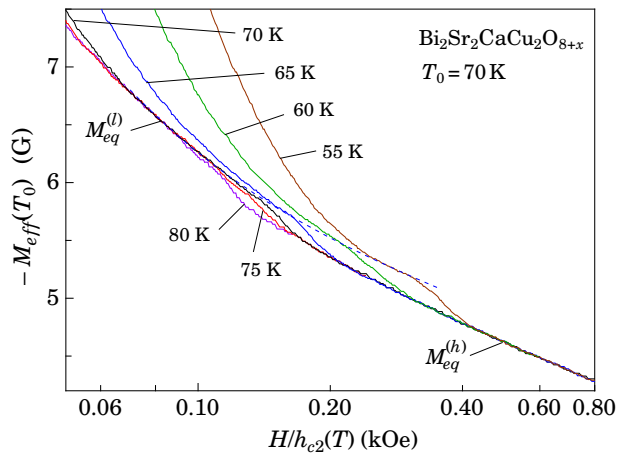}
  \caption{The $M_{eff}(H/h_{c2})$ curves for an optimally doped 
           Bi$_2$Sr$_2$CaCu$_2$O$_{8+x}$ sample.\cite{19} Only the 
           magnetizations measured in increasing fields are shown. The 
           dashed line indicates the most likely extrapolation of the 
           $M_{eq}^{(lp)}$ curve.}
 \end{center}
\end{figure}
\begin{figure}[h]
 \begin{center}
  \epsfxsize=0.95\columnwidth \epsfbox {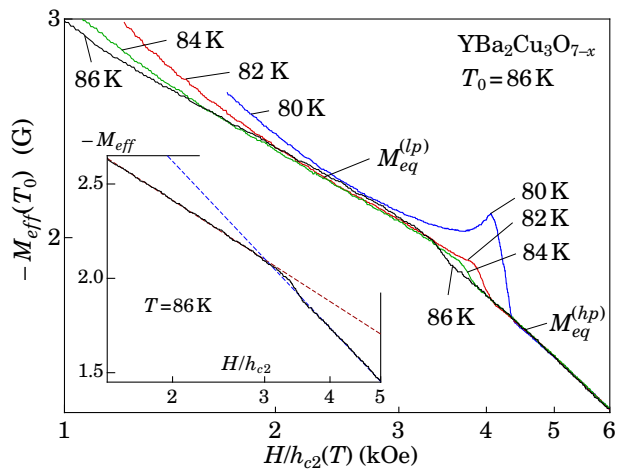}
  \caption{The $M_{eff}(H/h_{c2})$ curves for an optimally doped 
           YBa$_2$Cu$_3$O$_{7-x}$ sample, experimentally investigated 
           in Ref. \onlinecite{23}. Only the data collected in 
           increasing fields are shown. The inset shows 
           $M_{eff}(H/h_{c2})$ for $T=86$ K; the dashed straight lines 
           denote the linear extrapolations of $M_{eq}^{(lp)}(H)$ and 
           $M_{eq}^{(hp)}(H)$.}
 \end{center}
\end{figure}
First, we consider the experimental data for two similar 
(La$_{0.954}$Sr$_{0.046}$)$_2$CuO$_4$ (La-214) single crystals. 
The scaling results for these samples, calculated for $T_0 =$ 14 K, 
are shown in Fig. 1. In these two experiments, the magnetization is 
reversible down to magnetic fields well below the phase transition. 
The collapse of the data points measured at different temperatures is 
achieved with the same values of $h_{c2}(T)$ and $c_0(T)$ both below 
and above the transition, in complete agreement with the expectation 
outlined above. The results displayed in Fig. 1 may be interpreted as 
evidence for the existence of two different modifications of the mixed 
state with two different equilibrium $M_{eff}(H,T_0)$ curves above and 
below the transition. In the following we use $M_{eq}^{(lp)}$ and 
$M_{eq}^{(hp)}$ to distinguish the equilibrium $M_{eff}(H)$ curves at 
$T = T_0$ for the low- and the high-field modifications of the mixed 
state, respectively. As may be seen in Fig. 1, the low-field 
modification corresponds to somewhat higher values of the diamagnetic 
moment (smaller vortex density). The difference $\Delta M = {\left| 
{M_{eq}^{(lp)}} \right|} - {\left| {M_{eq}^{(hp)}} \right|}$ is 
positive. Analogous results for optimally doped 
Bi$_2$Sr$_2$CaCu$_2$O$_{8+x}$ (Bi-2212) and Y-123 samples are shown in 
Figs. 2 and 3, respectively. The behavior of the magnetization in 
the vicinity of the phase transition for these two samples is similar 
to that presented in Fig. 1. The $M_{eq}^{(lp)}$ and $M_{eq}^{(hp)}$ 
curves are readily identified, in spite of a pronounced peak effect 
observed for the Y-123 sample at the lowest temperature. The 
difference $\Delta M = {\left| {M_{eq}^{(lp)}} \right|} - {\left| 
{M_{eq}^{(hp)}} \right|}$ is again positive in both cases.

\begin{figure}[h]
 \begin{center}
  \epsfxsize=0.95\columnwidth \epsfbox {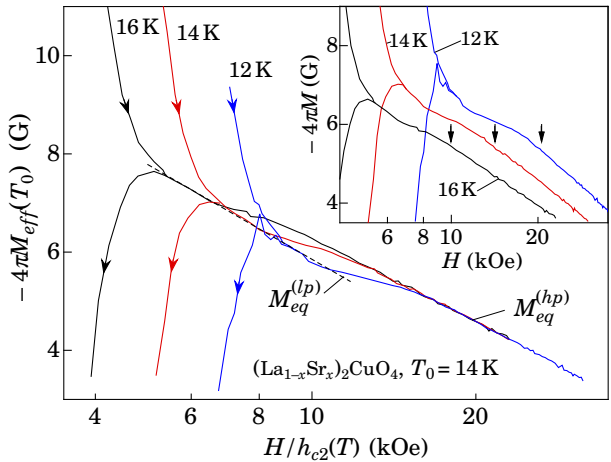}
  \caption{The $M_{eff}(H)$ curves for a La-214 sample studied in Ref. 
           \onlinecite{17}. The dashed line indicates 
           $M_{eq}^{(lp)}(H)$. The inset shows the original 
           magnetization data. The vertical arrows mark the phase 
           transition, as claimed in Ref. \onlinecite{17}.}
 \end{center}
\end{figure}
The $M_{eff}(H)$ curves in the transition region for another La-214 
sample, which are displayed in Fig. 4, are quite different from those 
shown in Figs. 1-3. The easily distinguishable $M_{eq}^{(lp)}(H)$ and 
$M_{eq}^{(hp)}(H)$ curves indicate that the difference $\Delta M = 
{\left| {M_{eq}^{(lp)}} \right|} - {\left| {M_{eq}^{(hp)}} \right|}$ is 
negative. The original magnetization data are shown in the inset of Fig. 
4; the only difference to Fig. 1 of Ref. \onlinecite{17} is that we use 
a log-scale for $H$. The arrows indicate the middle-points of the phase 
transitions at the corresponding temperatures, as claimed in Ref. 
\onlinecite{17}. Our plot reveals no evidence for phase transitions at 
these magnetic fields. We argue that the natural curvature of the $M(H)$ 
curves that were plotted on linear scales, emphasized by a real change 
of slopes at lower fields, lead to a questionable identification of the 
phase transitions.

\begin{figure}[h]
 \begin{center}
  \epsfxsize=0.95\columnwidth \epsfbox {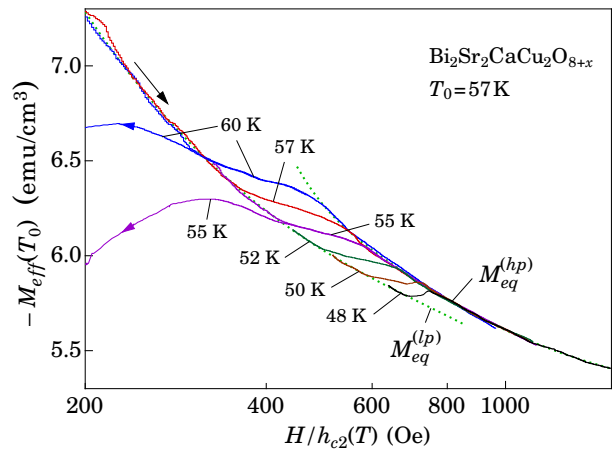}
  \caption{The $M_{eff}(H)$ curves for an overdoped Bi-2212 
           sample.\cite{18} The dotted lines indicate 
           $M_{eq}^{(lp)}(H)$ and $M_{eq}^{(hp)}(H)$. Because of rather 
           strong fluctuations of the magnetization measured in 
           decreasing fields at 48 K $\le T \le$ 52 K and T = 57 K 
           (see Fig. 2(a) of Ref. \onlinecite{18}), the corresponding 
           branches of the $M_{eff}(H/h_{c2})$ curves for these 
           temperature are not displayed in this plot.}
 \end{center}
\end{figure}
Finally we show the result of the scaling procedure for an overdoped 
Bi$_2$Sr$_2$CaCu$_2$O$_{8+x}$ sample in Fig. 5. Although only $M(H)$ 
data in magnetic fields above the transition were used for the 
evaluation of $h_{c2}(T)$ and $c_0(T)$, the magnetization curves 
measured in increasing magnetic fields collapse onto a single 
$M_{eff}(H)$ curve also below the irreversibility line and thus, as 
argued above, represent the equilibrium magnetization for $T = T_0$. 
\cite{14} The equilibrium $M(H)$ curves for the low and high field 
modifications of the mixed state are such that the difference $\Delta M = 
{\left| {M_{eq}^{(lp)}} \right|} - {\left| {M_{eq}^{(hp)}} \right|}$ in 
the transition region is again negative. The width of the transition from 
one modification of the mixed state to the other, which is quite large 
at higher temperatures, is considerably reduced with decreasing 
temperature. Also in this case we disagree with the interpretation of the 
magnetization curves given by the authors of the original publication. 
According to Ref. \onlinecite{18}, the difference $\Delta M$ changes its 
sign at $T \approx$ 50 K, while from Fig. 5 it is clear that $\Delta M$ 
is always negative and its absolute value monotonically decreases with 
decreasing temperature. 

The most unexpected result of our analysis is that the magnetization 
difference $\Delta M$ across the transition may adopt either sign. This 
result is difficult to reconcile with the vortex-lattice-melting 
hypothesis. In case of vortex lattice melting, the external magnetic 
field acts as pressure does in traditional solid-liquid melting 
transitions. Thermodynamics requires that the phase corresponding to the 
higher pressure must have a higher density, independent of whether this 
high-pressure phase is a liquid or a solid. In relation with the mixed 
state of type-II superconductors, the vortex liquid necessarily has to 
adopt a higher vortex density, i.e., the difference $\Delta M = {\left| 
{M_{eq}^{(lp)}} \right|} - {\left| {M_{eq}^{(hp)}} \right|}$ must always 
be positive. Since negative values of $\Delta M$ are identified for 
materials belonging to two different families of HTSC's, this can hardly 
be refuted as an accidental result. 

In the bulk of the existing literature, the first order transition in 
the mixed state of HTSC's is viewed as a melting transition in the 
system of vortices. In this scenario, the mixed state above the transition 
represents the vortex liquid, while the vortex solid below the transition 
is described as a lattice or Bragg glass of vortices, depending on the 
particular experimental conditions. Numerous experimental observations 
in the literature are in agreement with this interpretation. However, if 
the vortex lattice melting is indeed always responsible for the first 
order transition, all experimental results  must find their explanation 
from this point of view. Apparently, this is not the case. We see no way 
in which the negative values of $\Delta M$, clearly demonstrated in Figs. 
4 and 5, may be explained by invoking the vortex-lattice-melting 
hypothesis. 

We also note that in the case of the Y-123 sample, the slope 
$dM_{eq}/d(\ln H)$ in the low-field phase is substantially 
smaller than that in the high-field phase (see inset of Fig. 3). Because 
an order-disorder transition, such as the vortex lattice melting, cannot 
significantly change the field dependence of the sample magnetization, 
such a change of $dM_{eq}/d(\ln H)$ is not expected at a vortex 
lattice-melting-transition.

If there is a first order phase transition in the mixed state of HTSC's 
which is not related to vortex lattice melting, it must be of different 
origin. Below we present several possible scenarios. We do not argue in 
favor of one or the other possibility and we do not even claim that 
all these scenarios are indeed realistic. Of course, we cannot exclude 
other possibilities which are not considered here.

1. The change of the symmetry of the vortex lattice. It is well known 
that with increasing magnetic field a triangular vortex lattice may 
change to a square one via a first-order phase transition. In general, 
such a transition reduces the density of vortices. It is also possible, 
however, that a combination of this transition and a melting of the 
vortex lattice has to be considered. It is indeed possible that the 
position of the melting line $H_m(T)$ depends on the symmetry of the 
vortex lattice and two different melting lines $H_m^{(tr)}(T)$ and 
$H_m^{(sq)}(T)$ exist for the triangular and square configurations, 
respectively. If $H_m^{(sq)}(T) < H_{sym}(T) < H_m^{(tr)}(T)$, where 
$H_{sym}(T)$ denotes the symmetry transition, we have, with increasing 
field or temperature, a transition to a square vortex lattice which 
immediately melts down. In this case, there are two contributions to 
$\Delta M$ with opposite signs and the resulting value of $\Delta M$ 
may be positive or negative depending on the experimental conditions.

2. A magnetic field induced transition in the superconducting material 
which is not directly related to the mixed state but changes the 
superconducting parameters of the sample. It is difficult to imagine, 
however, that the temperature dependence of such a transition follows 
that of the first order phase transition in the mixed state of HTSC's.

3. It was suggested in Ref. \onlinecite{13} that in high magnetic 
fields, the mixed state of high-$\kappa$ superconductors is formed by 
superconducting filaments in a normal-state matrix, instead of 
Abrikosov vortices in a superconducting background. The transition to 
the vortex state with decreasing magnetic field occurs via a topological 
transition with all generic features of a first order phase transition.  
The sign of $\Delta M$ would depend on particular experimental 
conditions.

In conclusion, the most important result of our analysis is that the 
difference $\Delta M = {\left| {M_{eq}^{(lp)}} \right|} - {\left| 
{M_{eq}^{(hp)}} \right|}$ between the equilibrium magnetization 
curves below and above the transition in the mixed state of HTSC's may 
be negative as well as positive. The negative sign of $\Delta M$ is in 
serious conflict with the interpretation that this transition 
reflects the melting of the vortex lattice.

\end{document}